\section{Related Work and Background}


\subsection{Information Diffusion}

The study of opinion dynamics and information diffusion in social networks has a long history in the social, physical, and computational sciences~\cite{rapoport53-1, epstein96-1, axelrod97-1, castellano09-1, barrat08-1, leskovec06-1, leskovec09-1}. 
While usually referred to as `viral'~\cite{rapoport53-1}, the way in which information or rumors diffuse in a network has several important differences with respect to infections diseases. Rumors gradually acquire more credibility and appeal as more and more network neighbors acquire them. After some time, a threshold is crossed and the rumor becomes so widespread that it is considered as `common knowledge' within a community and hence, true.
In the case of information propagation in the real world as well as in the blogosphere, the problem is significantly complicated by the fact that the social network structure is unknown.  Without explicit linkage data investigators must rely on heuristics at the node level to infer the underlying network structure.  Gomez {\em et al.} propose an algorithm that can efficiently approximate linkage information based on the times at which specific URLs appear in a network of news sites~\cite{Gomez-Rodriguez:2010fk}. However, even in the case of the Twitter social network, where explicit follower/followee social relations exist, they are not all equally important~\cite{Huberman:2008uq}. Fortunately for our purposes, Twitter provides an explicit way to mark the diffusion of information in the form of {\em retweets.} This metadata tells us which links in the social network have actually played a role the diffusion of information.


Additionally, conversational aspects of social interaction in Twitter have recently been studied~\cite{boyd08-1, honeycutt08-1, goncalves10-1}. For example, Mendoza \textit{et al.} examined the reliability of retweeted information in the hours following the 2010 Chilean earthquake~\cite{Mendoza:2010kx}.  They found that false information is more likely to be questioned by users than reliable accounts of the event. Their work is distinct from our own in that it does not investigate the dynamics of misinformation propagation.  Finally, recent modeling work taking into account user behavior, user-user influence and resource virulence has been used to predict the spread of URLs through the Twitter social network~\cite{galuba-wosn10}.  



\subsection{Mining Microblog Data}

Several studies have demonstrated that information shared on Twitter has some intrinsic value, facilitating, e.g., predictions of box office success~\cite{asur10} and the results of political elections~\cite{tumasjan2010}.  Content has been further analyzed to study consumer reactions to specific brands~\cite{jansen09}, the use of tags to alter content~\cite{huang10}, its relation to headline news~\cite{Kwak10}, and on the factors that influence the probability of a meme to be retweeted~\cite{suh10}. Other authors have focused on how passive and active users influence the spreading paths~\cite{romero10}.  

Recent work has leveraged the collective behavior of Twitter users to gain insight into a number of diverse phenomena. Analysis of tweet content has shown that some correlation exists between the global mood of its users and important worldwide events~\cite{johan}, including stock market fluctuations~\cite{bollen-djia}. Similar techniques  have been applied to infer relationships between media events such as presidential debates and affective responses among social media users~\cite{Diakopoulos:2010}. Sankaranarayanan \textit{et al.} developed an automated breaking news detection system based on the linking behavior of Twitter users~\cite{Sankaranarayanan:2009zr}, while Heer and Boyd describe a system for visualizing and exploring the relationships between users in large-scale social media systems~\cite{Heer:2005ly}.  Driven by practical concerns, others have successfully approximated the epicenter of earthquakes in Japan by treating Twitter users as a geographically-distributed sensor network~\cite{Sakaki10earthquakes}.  



\subsection{Political Astroturf and Truthiness}

In the remainder of this paper we describe a system designed to detect astroturfing campaigns on Twitter. As an example of such a campaign, we turn to an illustrative case study documented by Metaxas and Mustafaraj, who describe a concerted, deceitful attempt to cause a specific URL to rise to prominence on Twitter through the use of a distributed network of nine fake user accounts~\cite{mustafaraj10-1}. In total these accounts produced 929 tweets over the course of 138 minutes, all of which included a link to a website smearing one of the candidates in the 2009 Massachusetts special election. The tweets injecting this meme mentioned users who had previously expressed interest in the Massachusetts special election, being prime candidates to act as rebroadcasters. By this the initiators sought not just to expose a finite audience to a specific URL, but to trigger an information cascade that would lend a sense of credibility and grassroots enthusiasm to a specific political message. Within hours, a substantial portion of the targeted users re\-tweeted the link, resulting in rapid spreading that was detected by Google's real-time search engine. This caused the URL in question to be promoted to the top of the Google results page for the query `{\tt martha coakley}' --- a so-called \emph{Twitter bomb.}  This case study demonstrates the ease with which a focused effort can initiate the viral spread of information on Twitter, and the serious consequences this can have.

Our work is related to the detection of spam in Twitter, which has been the subject of several recent studies. Grier {\em et al.} provide a general overview of spam on Twitter~\cite{Grier2010}, focusing on spam designed to cause users to click a specific URL. Grouping together tweets about the same URL into spam `campaigns,' they find a minimal amount of collusion between spammer accounts. Boyd {\em et al.} also analyze Twitter spam with respect to a particular meme~\cite{boyd-spam}. Using a hand-classified set of 300 tweets, they identify several differences between spam and good user accounts, including the frequency of tweets, age of accounts, and their respective periphery in the social graph. Benevenuto \textit{et al.}~\cite{benevenuto-ceas10} use content and user behavior attributes to train a machine learning apparatus to detect spam accounts. They build a classifier that achieves approximately 87\% accuracy in identifying spam tweets, and similar accuracy in detecting the spam accounts themselves.

The mass creation of accounts, the impersonation of users, and the posting of deceptive content are all behaviors that are likely common to both spam and political astroturfing. However, political astroturf is not exactly the same as spam. While the primary objective of a spammer is often to persuade users to click a link, someone interested in promoting an astroturf message wants to establish a false sense of group consensus about a particular idea. Related to this process is the fact that users are more likely to believe a message that they perceive as coming from several independent sources, or from an acquaintance~\cite{Jagatic:2007ys}.  Spam detection systems often focus on the content of a potential spam message --- for instance, to see if the message contains a certain link or set of tags. In detecting political astroturf, we focus on how the message is delivered rather than on its content. Further, many of the users involved in propagating a successfully astroturfed message may in fact be legitimate users, who are unwittingly complicit in the deception, having been themselves deceived by the original core of automated accounts. Thus, existing methods for detecting spam that focus on properties of user accounts, such as the number of URLs in tweets originating from that account or the interval between successive tweets, would be unsuccessful in finding such astroturfed memes.

In light of these characteristics of political astroturf, we need a definition that allows us to discriminate such falsely-propagated information from organically propagated information that originates at the real grassroots. We thus decided to borrow a term, \emph{truthy,} to describe political astroturf memes. The term was coined by comedian Stephen Colbert to describe something that a person claims to know based on emotion rather than evidence or facts. We can then define our task as the detection of truthy memes in the Twitter stream. Not every truthy meme will result in a viral information cascade like the one documented by Metaxas and Mustafaraj, but we wish to test the hypothesis that the initial stages exhibit common signatures that can help us identify this behavior.